\documentclass[reprint,amsmath,amssymb, aps, prb, showkeys, superscriptaddress]{revtex4-1}

\usepackage{graphicx}% Include figure files
\usepackage{dcolumn}% Align table columns on decimal point
\usepackage{bm}% bold math
\usepackage{color}
\usepackage{natbib}
\usepackage{appendix}
\usepackage{dsfont}

\usepackage{appendix}

\begin{document}

\title{Anomalous transport in periodic photonic chains with designed loss}

\author{I. Peshko} 
\affiliation{B.I.Stepanov Institute of Physics, NAS of Belarus, Nezavisimosti ave. 68, 220072 Minsk, Belarus}

\author{G. Ya. Slepyan}
\affiliation{School of Electrical Engineering, Tel Aviv University, Tel Aviv 69978, Israel}

\author{D.Mogilevtsev}
\affiliation{B.I.Stepanov Institute of Physics, NAS of Belarus, Nezavisimosti ave. 68, 220072 Minsk, Belarus}

\date{September 2022}

\begin{abstract}
Here we show that a coherent random walk in a perfectly periodic chain of  bosonic modes with designed loss  can exhibit a variety of different anomalous transfer regimes in dependence on the initial state of the chain. In particular, for any given finite initial time-interval there is a set of initial states leading to a hyperballistic transport regime. Also, there are initial states allowing one to achieve a subdiffusive regime or even localization for a given  time-interval, or change an asymptotic long-time diffusion rate. We show how these anomalous transport regimes can be practically realized in a laser-written network of single-mode waveguides in balk glass or how a planar system of coupled single-mode waveguides can be realized with an integrated photonic platform. 
\end{abstract}

\maketitle

\section{Introduction} 

Due to recent development of experimental techniques allowing for creating  involved networks of high-quality optical waveguides \cite{davis1996,Szameit_2010},  systems of coupled single-mode waveguides have become a powerful tool for emulating a plethora of effects reaching even up to the cosmological ones \cite{2019Natur,sheng,PhysRevLett.109.110401}. Systems of evanescently coupled single-mode waveguides in a balk dielectric allow one to create perfectly periodic flat-band structures where localization is possible for certain states \cite{PhysRevLett.114.245503,PhysRevLett.114.245504,wang2020}, demonstrate Aharonov-Bohm caging \cite{longhi2014,PhysRevLett.121.075502} and different kinds of topological insulation \cite{kremer,PhysRevLett.125.213901,rechtsman}, emulate synthetic dimensions \cite{Yuan18} and even the gauge field of a cosmic string \cite{sheng}.

Adding controlled designed loss to such waveguides networks opened a number of novel interesting possibilities. They were deemed interesting even to such an extent as to produce a title "Loss leads the way to utopia" in a very recent issue of Nature Photonics \cite{loss}. Designed loss in a waveguide network can induce a plethora of effects such as  nonreciprocal propagation and directional amplification \cite{PhysRevX.5.021025,arenz,Huang2021,myarxiv2022},  non-Hermitian skin effects \cite{longhi2020},    deterministic and conditional nonclassical state generation and protection \cite{PhysRevLett.86.4988,PhysRevApplied.12.064051}, and robust multiple-beam splitting  and optical routing  \cite{Mukherjee2017DissipativelyCW}. 

There are quite feasible experimental possibilities of creating large networks of coupled waveguides with designed loss (with loss realized by coupling to, for example,  as "waveguide tails" \cite{bigger,Mukherjee2017DissipativelyCW} or through "wavy waveguides" \cite{eichel,grafe} laser-written in balk glass,  or just planar semiconductor structures with metal added on the tops of waveguides \cite{myarxiv2022}). However, there are not many studies that discuss state transport between the waveguides, especially in the regime when the designed loss dominates the dynamics (realizing so called "dissipative coupling"). To our knowledge, still there are no works  describing coherent effects of this transport stemming from  delocalized excitation of dissipatively coupled waveguides, i.e.,  of exciting initially several waveguides in the chain.   From the other side, there are already clear hints about such a coherent transport being able to reveal interesting unexpected features. Recently is was shown that decoherence and loss can turn a coherent walk into a classical one and switch the ballistic regime to the diffusive one after some interaction time even in perfectly periodic lattices \cite{kendon,romanelli,eichel,loss1}.  

Already for quite a long time, coherent walks have been an intensively studied field of research \cite{aharon93,kempe,kendon,venegas}. Their time-continuous version represents a general scheme of quantum state transport through a network of coupled quantum systems and can be used for consideration of numerous transport phenomena in solid-state physics, biophysics, photonics, etc. \cite{farhi,kempe,MULKEN201137,PhysRevLett.100.170506,PhysRevLett.99.090601,reben,PhysRevE.93.032407,grafe}. Typically, in the absence of decoherence and loss, quantum walks demonstrate ballistic spatial spread in contrast with classical random walks demonstrating diffusive spread \cite{kempe}. It is customarily considered that alteration of spatial variance behavior can be achieved by spatial or temporal perturbation of the lattice order or by nonlinear effects. For example, it was shown that one can achieve hyperballistic or subdiffusive transport regimes, or even localization, by randomly perturbing the network and making the structure nonhomogeneous, or via nonlinearity, or controlled time-dependent operations, or by subjecting the network to non-Markovian noise, making the probabilities of jumps between the network systems time-dependent \cite{liad,levi2012,PhysRevLett.108.070603,PhysRevLett.100.094101,PhysRevA.105.042210,PhysRevLett.106.180403,PhysRevE.70.045101}.

\begin{figure}[htb]
\begin{center}
\includegraphics[width=\linewidth]{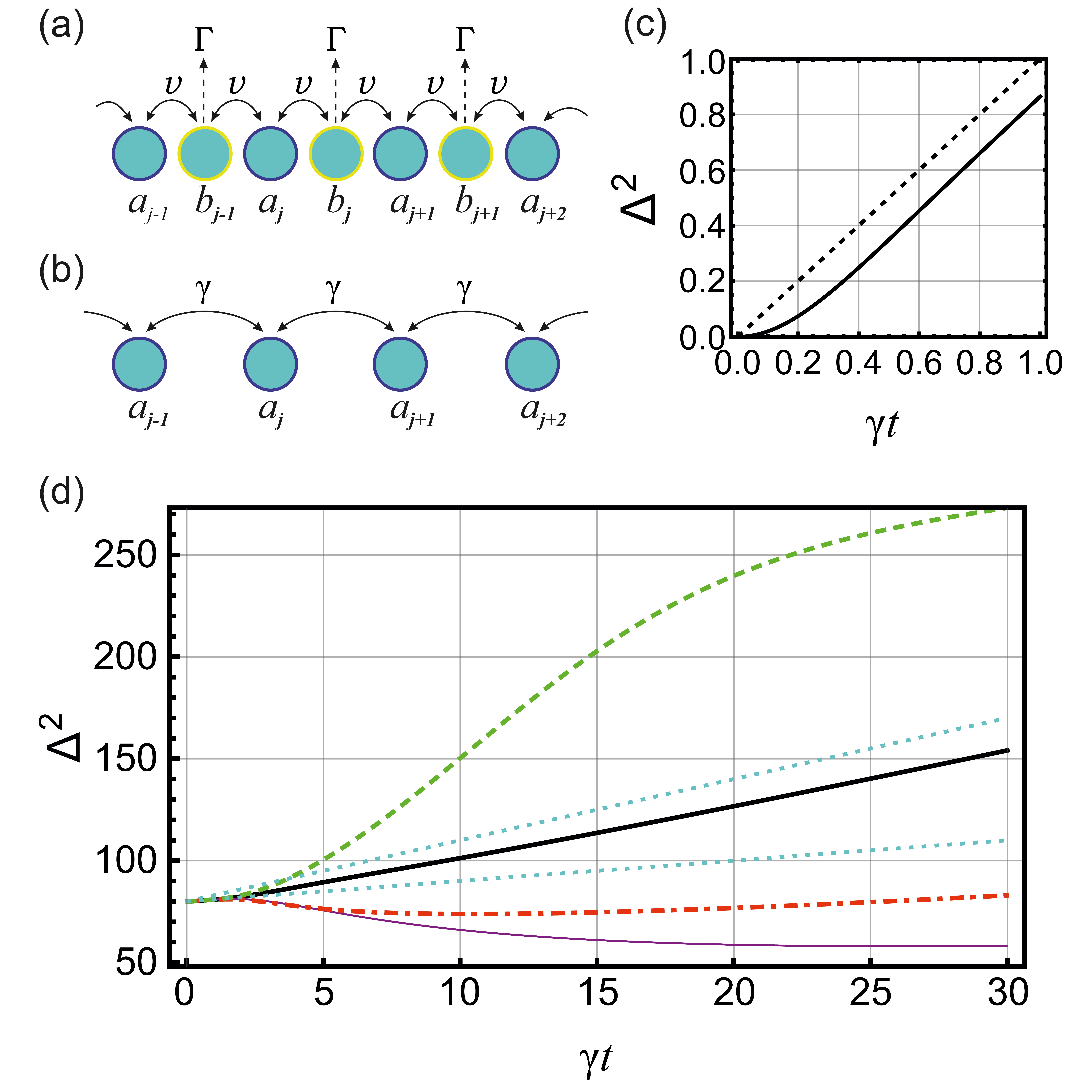}
\end{center}
\caption{ (a) The scheme of a bosonic mode chain with designed loss in waveguides $b_j$  described by the matrix equation (\ref{lind3}). (b) The scheme of the dissipatively coupled tight-binding chain of bosonic modes described by Eq.(\ref{lind2}). (c) The variance  for the single initial excited mode   with the number $n_0=150$ in the chain with 300 waveguides. The dotted line corresponds to the linear dependence $\gamma t$. 
(d) The variance for the initial state given by Eq.(\ref{inst}) for four initially excited modes with the numbers $k=n_0-8,n_0,n_0+8,n_0+16$ for $n_0=150$ in the chain with 300 waveguides. Thick solid, thin solid, dashed and dash-dotted lines correspond to the following state signatures:  $\{+,+,-,-\}$, $\{+,-,-,-\}$, $\{-,+,-,+\}$, $\{+,+,+,+\}$. Upper and lower dotted lines correspond to the linear dependencies $3\gamma t$ and $\gamma t$. }
\label{fig1}
\end{figure}

Here we demonstrate that  coherent interference effects in a perfectly periodic linear and unperturbed waveguide network with designed loss allow one to obtain quite a wide spectrum of transport regimes. In particular, for an arbitrary finite time interval one can achieve superballistic  growth of a spatial variance of the field distribution in the waveguide chain, or subdiffusive and even localization behavior for a timescale much exceeding the typical timescale of excitation exchange between neighboring waveguides. For all the considered cases, a long-time  asymptotic behavior always remains diffusive. However, the rate of diffusion depends on the initial state. 

The paper is structured in the following way. In Section II we describe the model of the next-neighbor coupled chain of bosonic modes realized by evanescently coupled single-mode waveguides, derive the master equation in the dissipative coupling limit, obtain the system of equations for the energy transfer across the chain and discuss its solution. In Section III, we identify the stages of transport and show how the different transport regimes arise during the interference stage  in dependence on the initial state. In particular, we show how  superballistic transport and localization can appear during times much exceeding typical transient timescales for the considered modal structure. Also, we show how these regimes scale with delocalization  of the initial state.  In Section IV, we discuss the character of interference leading to different transport regimes and derive a simple model able to provide for a qualitative description of the regimes and their scaling behavior. In Section V, we discuss practicalities of possible realization of our regimes in a waveguide system with designed loss and illustrate how the dissipative coupling limit can be approached in practice. 

\section{Model} 

In our work we discuss a one-dimensional (1D) chain of next-neighbor linearly coupled bosonic modes with every second mode subjected to additional designed loss (see Fig~\ref{fig1} (a)). This scheme can be described by the following master equation for the total density matrix of the system
\begin{eqnarray}
\label{lind3}
\frac{d}{dt}\rho_{tot}=-\frac{i}{\hbar}[H,\rho_{tot}]+\\
\nonumber
(\Gamma+\gamma_{bulk})\sum\limits_{\forall j}\Bigl(2b_j\rho_{tot} b_j^{\dagger}-
\rho_{tot} b_j^{\dagger}b_j-b_j^{\dagger}b_j\rho_{tot}\Bigr)+ \\
\nonumber
\gamma_{bulk}\sum\limits_{\forall j}\Bigl(2a_j\rho_{tot} a_j^{\dagger}-
\rho_{tot} a_j^{\dagger}a_j-a_j^{\dagger}a_j\rho_{tot}\Bigr),
\end{eqnarray}
where the modal interaction Hamiltonian is 
\begin{eqnarray}
H=\sum\limits_{\forall j} \hbar v\left( b_j^{\dagger}(a_{j}+a_{j+1})+(a_{j}^{\dagger}+a_{j+1}^{\dagger})b_j\right).
    \label{ham1}
\end{eqnarray}
The operators $b_j$ and $b_{j}^{\dagger}$ are the annihilation and creation operators for modes of the waveguides with the additional designed loss with the rate $\Gamma$. The  annihilation and creation operators $a_j$ and $a_{j}^{\dagger}$ describe modes in  waveguides with only the passive loss unavoidably occurring in realistic waveguide systems. The rate of this loss is $\gamma_{bulk}$. The interaction constant $v$ describes the beating rate between the waveguides. 

In Sec. V, we discuss in more details practical realizations of the scheme (\ref{lind3}) as the network of next-neighbor evanescently coupled single-mode waveguides. In particular, we show that it is quite feasible to create waveguide chains with acceptably low homogeneous passive loss (actually, of about 1 dB/cm or lower) and sufficiently high designed loss $\Gamma$. Thus the whole range of the effects discussed below is feasible for practical observation in systems of waveguides laser-written in balk glass or in planar semiconductor structures  \cite{bigger,Mukherjee2017DissipativelyCW,eichel,grafe,myarxiv2022}.

\subsection{Dissipative coupling limit}

Here we concentrate our attention mostly on the case when the loss rate $\Gamma$ is large. We assume  that the state of modes $b_j$ is reduced to the vacuum on the timescale much shorter than the time-scale of the modes $a_j$ dynamics. Thus, the lossy modes $b_j$ can be adiabatically eliminated. The modes $a_j$ interact only through their coupling to the common loss reservoir.   Such "dissipatively coupled" systems have a number of remarkable properties that are useful for photonic applications. They "termalize"   coherence (i.e., off-diagonal elements of the density matrix in the energy basis \cite{Mogilevtsev_2015} or amplitudes of the modal coherent states \cite{Mukherjee2017DissipativelyCW}). So, they were considered as a basis for so-called "coherent diffusive photonics" for building optical equalizers/beam splitters, distributors and non-reciprocal optical circuits \cite{Mukherjee2017DissipativelyCW,myarxiv2022}. Nonlinear dissipatively coupled waveguides can be used  to generate and preserve nonclassical states \cite{PhysRevLett.86.4988,PhysRevX.5.021025,Mogilevtsev:10,PhysRevApplied.12.064051}
By adding  unitary coupling between modes and adjusting dissipative coupling between them, one can achieve unidirectional field transfer along the chain and create topologically nontrivial bound states \cite{PhysRevX.5.021025,arenz}.  

In the limit of large loss $v/\Gamma\rightarrow 0$, the lossy modes $b_j$ can be adiabatically eliminated from the master equation (\ref{lind3}). The result is the following master equation describing the system of dissipatively coupled bosonic modes $a_j$ schematically shown in Fig.\ref{fig1} (b) (for the detailed derivation  see, for example, Ref.\cite{Mogilevtsev:10}). 
\begin{eqnarray}
\nonumber
    \frac{d}{d t}\rho=\sum\limits_{\forall j}({\bar\gamma}+\gamma_{bulk})(2L_j\rho L_j^{\dagger}-\rho L_j^{\dagger}L_j-L_j^{\dagger}L_j\rho)+ \\
    \sum\limits_{\forall j}\gamma_{bulk}(2a_j\rho a_j^{\dagger}-\rho a_j^{\dagger}a_j-a_j^{\dagger}a_j\rho),
\label{lind2}
\end{eqnarray}
where ${\bar\gamma}=v^2/(\Gamma+\gamma_{bulk})$ is the dissipation rate and the Lindblad operators $L_j=a_j+a_{j+1}$ describe symmetric next-neighbor dissipative coupling between $j$th and $j+1$th modes with the annihilation operators $a_j$ and $a_{j+1}$. 

\subsection{Transport description} 

To discuss energy transport across the system of dissipatively coupled waveguides, we derive from the master equation (\ref{lind2})  the following equation for the first-order correlation functions:
\begin{equation}
\frac{1}{\gamma}\frac{d}{dt}g_{j,k}=-4g_{j,k}-g_{j+1,k}-g_{j-1,k}-g_{j,k+1}-g_{j,k-1},
\label{eqg1}
\end{equation}
where  $\gamma={\bar\gamma}+\gamma_{bulk}$. 
We account for the passive loss of waveguides by introducing  $g_{j,k}(t)=\langle a^{\dagger}_j(t)a_k(t)\rangle \exp\{2\gamma_{bulk}t\}$. 

To describe the transport, we introduce the normalized population distribution 
\begin{equation}
p_j=\frac{g_{j,j}}{p_{tot}}, \quad p_{tot}= \sum\limits_{\forall j}g_{j,j}.
\label{p}    
\end{equation}
Further we assume the case when the initially excited modes are far from the edges of the chain. We also take that during the timeinterval of our interest excitation is still far from the edges. We consider the dynamics of the excitation spread through the modes described by the spatial variance $\Delta^2$:
\begin{equation}
\Delta^2(t)=\sum\limits_{\forall j} (j-{\bar x}(t))^2p_j(t), \quad {\bar x}(t)=\sum\limits_{\forall j} jp_j(t), 
\label{var1}
\end{equation}
where ${\bar x}$  is the average time-dependent excitation shift. Notice that the variance $\Delta^2$ does not depend on $\gamma_{bulk}$ in the dissipative coupling limit when the designed loss much exceed the passive one, i.e., $\Gamma\gg\gamma_{bulk}$. 

Notice that by a trivial sign change, 
\[g_{j,k}\rightarrow (-1)^{j+k}g_{j,k}\], 
Eq.(\ref{eqg1}) transforms to the equation formally coinciding with the one describing a 2D classical continuous random walk \cite{kempe,alma990023849800205776}. So, for the chain of finite length described by Eq.(\ref{eqg1}), the exact analytic solution can be easily found \cite{alma990023849800205776,Mogilevtsev_2015} for any initial set of $g_{j,k}$.  One also has 
here a "coherence preservation": $\sum\limits_{\forall j}(-1)^{j+k}g_{j,k}(t)=const$. For negligible passive loss such a coherence preservation can lead to stationary entangled states despite the dissipative character of the dynamics \cite{Mogilevtsev_2015}.  It is worth emphasizing that, in contrast with the classical random walk,  $g_{j,k}$ are complex-value functions.

\begin{figure}[htb]
\begin{center}
\includegraphics[width=\linewidth]{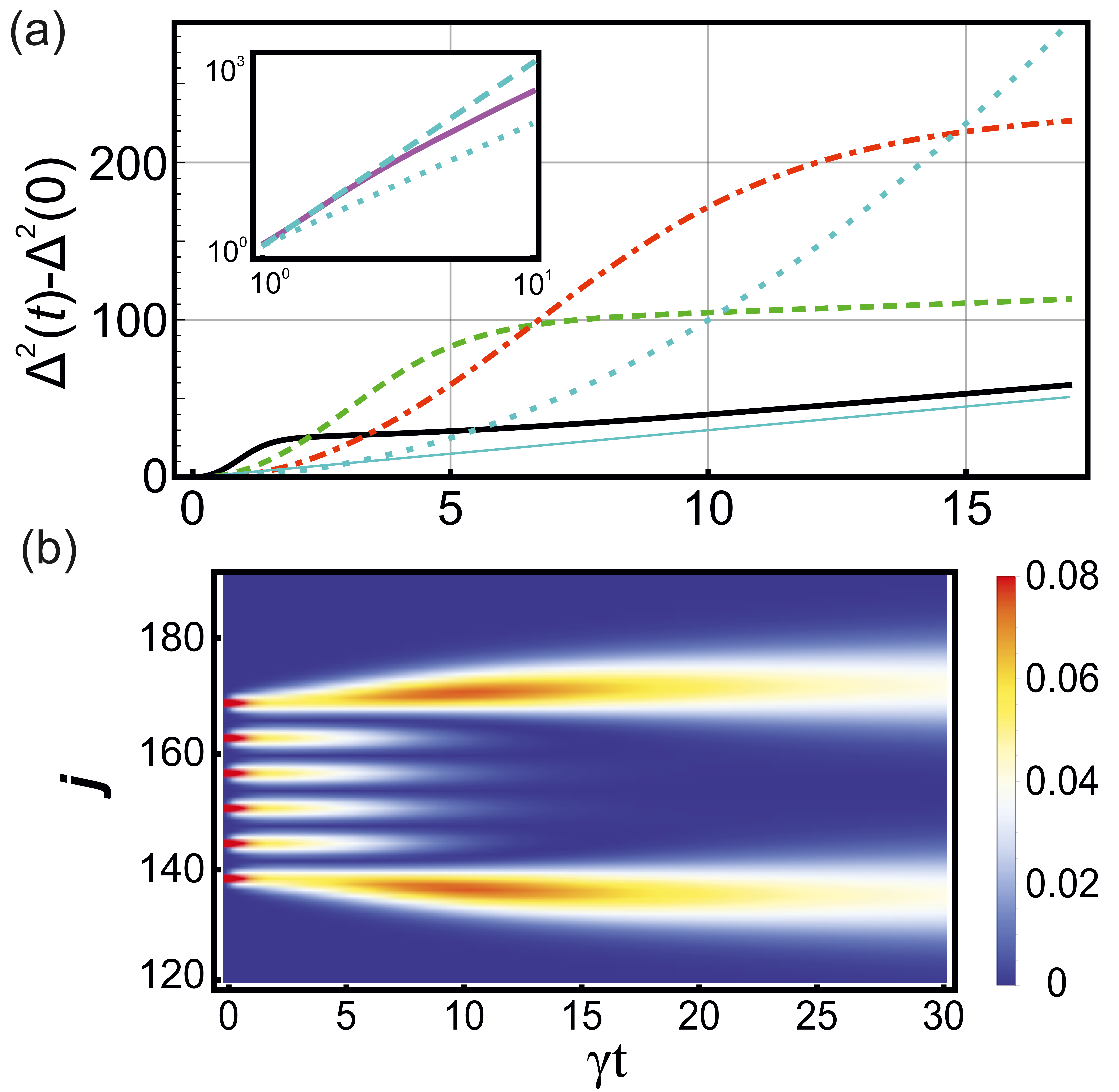}
\end{center}
\caption{(a) Illustration of the super-ballistic regime scaling for the initial state given by Eq.(\ref{supes1}). Thick solid, dashed and dash-dotted lines correspond to  $m=1,2,3$ and the function $\Delta^2(t)-\Delta^2(0)$. The thin solid line corresponds to the diffusive regime $\Delta^2(t)-\Delta^2(0)=3\gamma t$. The dotted line shows the ballistic regime $\Delta^2(t)-\Delta^2(0)=\gamma^2 t^2$. The inset shows log-log plots of the variance given by Eq.(\ref{supes2}). In the inset, the dashed and dotted lines show the super-ballistic and ballistic regimes given by $\Delta^2(t)-\Delta^2(0)=1.4(\gamma t)^n$, $n=3,2$. (b) The distribution of $p_j(t)$ for the initial state corresponding to $m=3$. Other parameters are the same as those for Fig.~\ref{fig1}(b).}
\label{fig2}
\end{figure}

\section{Transport} 

Here we discuss the appearance of different transport regimes as a consequence of modal interference, and we show how the initial state of the waveguide chain affects both the length and the character of the interference stage in dynamics. We demonstrate interference effects similarly to how it is usually done for classical antenna arrays showing how respective phases and spatial placement of excitations in a chain affect the field pattern. So, we consider a case when only some sets of modes  $\{k\}$ are initially excited in coherent states with equal absolute amplitudes but with different combinations of possible $\pi$-differences in phase, whereas all the other modes are initially in the vacuum state:
\begin{equation}
\rho(0)=|\psi\rangle\langle\psi|, \quad |\psi\rangle=\prod_{\forall  k}|\delta_{k}\alpha\rangle_{k}\prod_{\forall j\neq k}|0\rangle_{j},
    \label{inst}
\end{equation}
where $\delta_{k}=\pm 1$.

\subsection{Stages and regimes}

It needs to be mentioned that the system shown in Fig.~\ref{fig1}(a) was considered in Ref.\cite{eichel}. There only a single initially excited waveguide was considered outside of the regime of purely dissipative coupling (i.e., not in the limit  $v/\Gamma\rightarrow 0$). In Ref.\cite{eichel} it was found that initially, for time intervals less than  $\Gamma/v^2$, the transport is typical for the unitary coupled chain. It is ballistic. However, for times exceeding $\Gamma/v^2$ the transport acquires diffusive character with the linearly spreading variance. 

This result can be illustrated with the following simple estimation. For just a pair of dissipatively coupled modes, $a_{1,2}$,  with just one mode initially excited with the unit amplitude, one gets from Eq.(\ref{eqg1}) that 
\begin{equation}
 \label{two}   
p_{1,2}(t)=\frac{1}{2}\frac{(1\pm e^{-\gamma t})^2}{1+e^{-2\gamma t}}. 
\end{equation}
So, energy exchange between modes occurs on the timescale $~1/\gamma$. In Fig.~\ref{fig1}(c) the variance $\Delta^2(t)$ given by Eq.(\ref{var1}) is shown when just a single waveguide is initially excited.  There one can see that indeed for times exceeding $~1/\gamma$ the transport becomes diffusive (for comparison, the dotted line shows the dependence $\Delta^2(t)=\gamma t$). Curiously, as Eqs.(\ref{var1},\ref{two}) show, even in a simple two-waveguide model for small times, $\gamma t\ll 1$, the variance  has a ballistic character, $\Delta^2(t)\propto t^2$.   

Our main message is that adjusting phases, numbers and spatial separation of initially excited waveguides, due to interference, one can obtain a wide variety of transport regimes during a time interval (interference stage) much exceeding  $~1/\gamma$. 

Fig.~\ref{fig1}(d) shows an example of this variety obtained changing on $\pi$ phases of only four initially excited waveguides; i.e., we took  $k=n_0-2m,n_0,n_0+2m,n_0+4m$ in Eq.(\ref{inst}), where $n_0$ is the number of the waveguide in the middle of the chain, $m=4$, and $\alpha=1$. We chose this state to show to the maximal extent  both the interference stage and the asymptotic stage for all the considered cases.

 First of all, the result of the signature choice, $\mathrm{sign}[{\delta_{1,2,3,4}}]=\{+,+,-,-\}$, shown by the thick solid curve in Fig.~\ref{fig1}(d) gives an intuitively expected linear diffusion for times exceeding a typical interaction time scale ($\gamma t >1$), but with an unexpected rate: $3\gamma$ instead  of classically expected $\gamma$ (upper dotted line shows the dependence $\Delta^2(t)=3\gamma t$). 
 
 The signature $\{+,+,+,+\}$ gives an expected linear diffusion with the rate $\gamma$ shown by the dash-dotted curve in Fig.~\ref{fig1}(d) just like it was for the case of the single excited waveguide shown in Fig.~\ref{fig1}(c) (for comparison, the lower dotted line shows the dependence $\Delta^2(t)=\gamma t$). However,  this stage occurs only after some period of the variance decreasing.  The duration of this stage significantly exceeds the inverse rate $\gamma^{-1}$.  
 
 The signature $\{+,-,-,-\}$ gives only the localization during the considered interaction time, i.e., the variance  decreases  (as shown by the thin solid line in Fig.~\ref{fig1}(d)). 
 
 Finally, the signature  $\{-,+,-,+\}$ leads to a faster than diffusive transport  (the dashed curve in Fig.(\ref{fig1})(d)).

One can see that the choice of phases $\delta_j$ indeed can drastically affect the  dynamics of the variance $\Delta^2(t)$. The character of the transport  crucially depends on whether the initial state amplitudes sum to zero or not. We discuss reasons for this in the next section.

\begin{figure}[htb]
\begin{center}
\includegraphics[width=\linewidth]{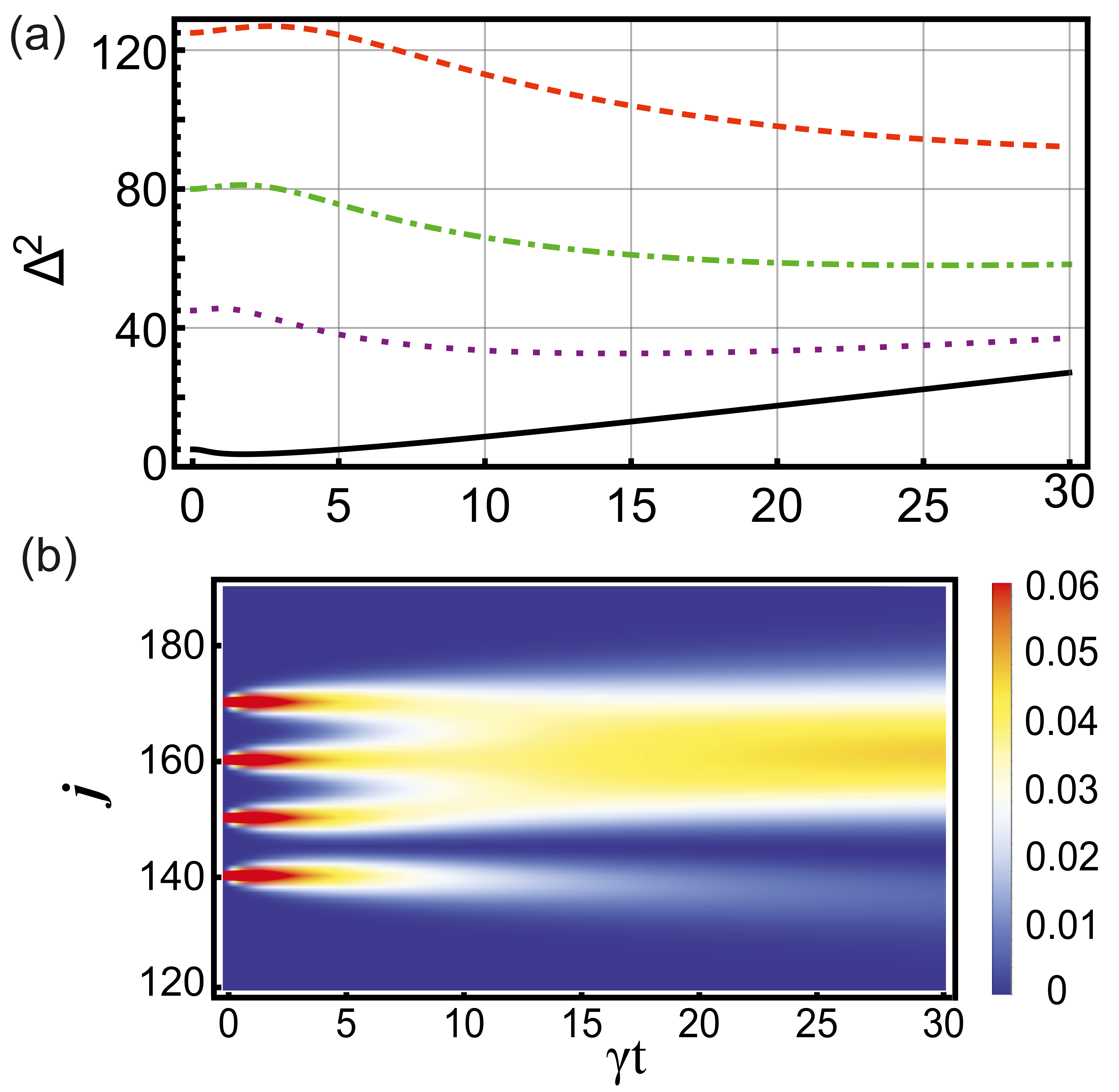}
\end{center}
\caption{ (a). Scaling of the localization regime for the initial state (\ref{inst}) with four initially excited modes with numbers $k=n_0-2m,n_0,n_0+2m,n_0+4m$ and the signature $\{+,-,-,-\}$. Solid, dotted, dash-dotted and dashed lines correspond to $m=1,3,4,5$. 
%(b) The probabilities $p_j$ distribution for the initial state corresponding to $m=5$ for $\gamma t=0$ (separate dark bars) and for $\gamma t=30$ (extended grey distribution) . 
(b) The distribution of $p_j(t)$ for the initial state corresponding to $m=5$. Other parameters are the same as those for Fig.(\ref{fig1})(d).}
\label{fig3}
\end{figure}

\subsection{Superballistic regime and scaling} 

Now let us show that the faster-than-diffusive regime exemplified by the dashed curve in Fig. (\ref{fig1})(d)) can be indeed superballistic.  Also, it can be scaled up by increasing the spatial extent of the initial state.  This case is illustrated in the main panel of Fig.~\ref{fig2}(a). Here the variance dynamics is given for the state (\ref{inst}) with six initially excited modes with the numbers 
\begin{equation}
k=n_0-4m,n_0-2m,n_0,n_0+2m,n_0+4m,n_0+6m
\label{supes1}
\end{equation}
with the amplitude $\alpha=1$ and the signature $\{-,+,-,+,-,+\}$. Thick solid, dashed and dash-dotted lines correspond to the initial state spatially extended with  $m=1,2,3$. For comparison, the diffusive regime $\Delta^2(t)=3\gamma t$ is shown by the thin solid line, and the ballistic regime $\Delta^2(t)=\gamma^2 t^2$ is shown by the dotted line; for better comparison the values of $\Delta^2(t)-\Delta^2(0)$ were shown for each case.  One can see how the superballistic region extends toward longer interaction times with increasing of the localization region of the initial state.

Remarkably, for the initial state (\ref{supes1}) the spatial distribution of the population shown in Fig.~\ref{fig2}(b) exhibits structuring instead of intuitively expected spreading and homogenization.  Only the population of the modes at the edges of the excited state survives, whereas inner modes are quickly depopulated.  In the next section we explain this seemingly counterintuitive behavior.

By choosing an initial state one can also the transition from the superballistic to the ballistic stage, with both of them extending for periods of time much exceeding $1/\gamma$. An example of such a state is the initial state (\ref{inst}) with 12 initially excited modes with $\alpha=1$, the numbers 
\begin{equation}
k=n_0+jm, \quad j=-5,-4,\ldots 5,6, \quad m=7,
\label{supes2}
\end{equation}
and the signature $\{(-1)^j\}$.  The inset in Fig.~\ref{fig2} plotted in log-log axes shows that there is indeed superballistic dynamics; it can have an extended, over several $1/\gamma$ intervals, stage when $\Delta^2(t)-\Delta^2(0)\propto (\gamma t)^3$. The solid line in the inset corresponds to the initial state (\ref{supes2}), the dotted line  in the inset corresponds to $\Delta^2(t)-\Delta^2(0)\approx 1.4(\gamma t)^2$, and the dashed line corresponds to $\Delta^2(t)-\Delta^2(0)\approx 1.4(\gamma t)^3$.  As shown in the inset, the transport corresponding to the initial state (\ref{supes2}) goes from super-ballistic (when the variance is cubic on the propagation time) to the superballistic regime (when the variance is proportional to the squared propagation time). However, it should be noticed that eventually for large interacting times (for the case of the initial state (\ref{supes2}), it is approximately for $\gamma t > 20$),  the transport is diffusive. 
\begin{figure}[htb]
\begin{center}
\includegraphics[width=\linewidth]{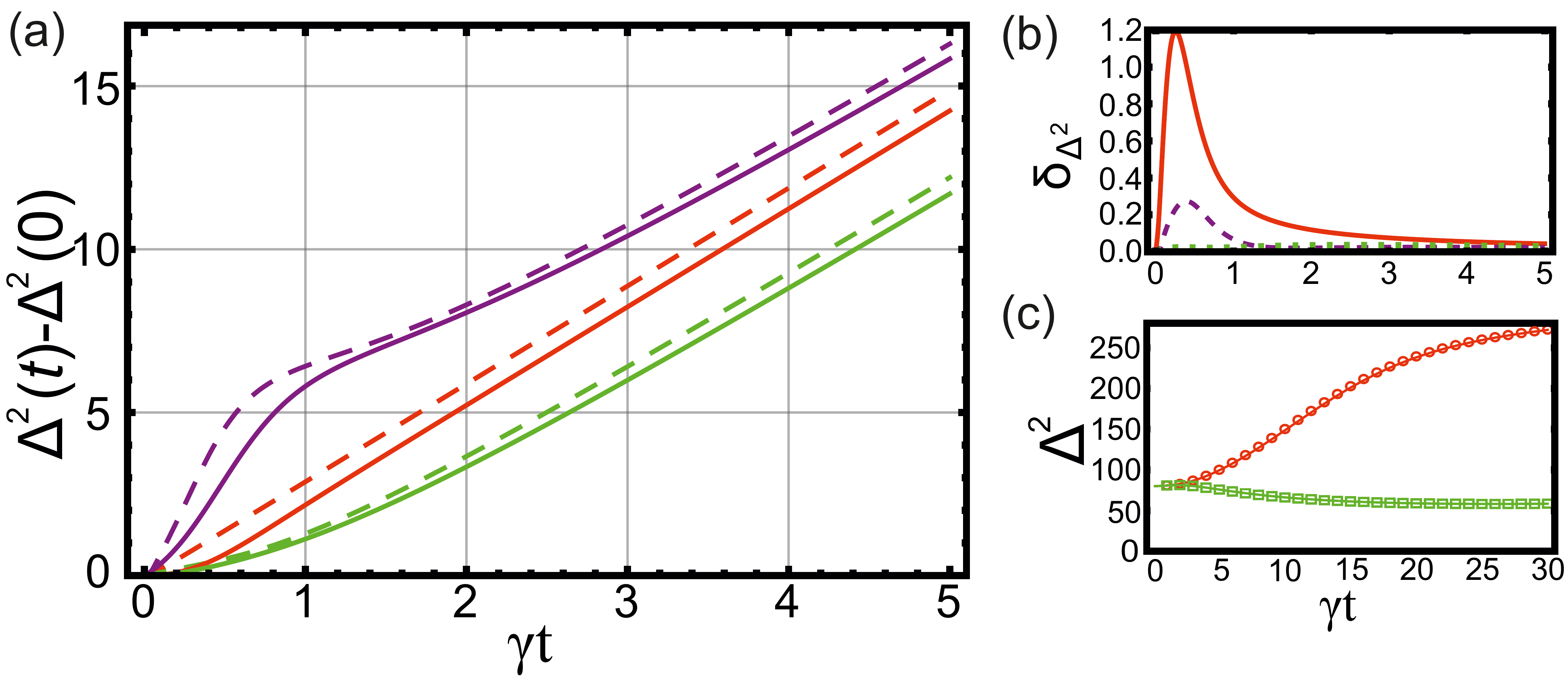}
\end{center}
\caption{ Illustration of the continuous approximation validity. Panel (a) shows the relative variance $\Delta^2(t)-\Delta^2(0)$. Here the solid lines correspond to the exact solution of Eq.(\ref{lind2}); the dashed lines correspond to the approximation given by Eqs.(\ref{varcon},\ref{varcoh}).  The middle solid and dashed lines correspond to two initially excited modes, $n_0-1$ and $n_0$, with the signature $\{+,-\}$. The lower solid and dashed lines correspond to two initially excited modes, $n_0-2$ and $n_0+3$, with the signature $\{+,-\}$.  The upper solid and dashed lines correspond  to four initially excited modes, $n_0-2,n_0,n_0+1,n_0+3$, with the signature $\{+,-,+,-\}$. The panel (b) shows a relative difference of variance for the exact and approximate solutions, $\delta_{\Delta^2}=|\Delta^2_{exact}(t)-\Delta^2_{approx}(t)|/\Delta^2_{exact}(t)$. The  solid line corresponds to two initially excited modes, $n_0-1$ and $n_0$, with the signature $\{+,-\}$. The dotted line corresponds to two initially excited modes, $n_0-2$ and $n_0+3$, with the signature $\{+,-\}$.  The dashed line corresponds  to four initially excited modes, $n_0-2,n_0,n_0+1,n_0+3$, with the signature $\{+,-,+,-\}$. Panel (c) shows how the continuous approximation captures the transport dynamics for the initial states considered in Fig.~\Ref{fig1}(c) and demonstrates the variance for the initial state given by Eq.(\ref{inst}) for the four initially excited modes with the numbers $k=n_0-8,n_0,n_0+8,n_0+16$. Solid and dashed lines correspond to the exact solutions with the  following state signatures:  $\{+,-,+,-\}$, $\{+,-,-,-\}$; circles and squares show corresponding approximate solutions.   For all panels, $n_0=150$ in the chain with 300 waveguides.  }
\label{newfig}
\end{figure}

\subsection{Localization and scaling} 

Now let us demonstrate that the localization evidenced by the thin solid line in Fig.~(\ref{fig1})(d)) can also be scaled; i.e., by choosing the initial state one can extend the region of localization to an arbitrary degree. We illustrate this situation with the same initial state (\ref{inst}) as was used for Fig.~(\ref{fig1})(d)); i.e., we have four initially excited modes with numbers $k=n_0-2m,n_0,n_0+2m,n_0+4m$ and the signature $\{+,-,-,-\}$.  Solid, dotted, dash-dotted and dashed lines correspond to increased spatial extent of the initial state with  $m=1,3,4,5$. One can see how the region of variance diminishing is extended toward the larger interaction times with extending the spatial localization of the initial state. One can also see in Fig.~(\ref{fig3})(b) that for the case of $m=5$ the distribution of the probabilities $p_j(t)$ remains indeed  well localized during all the interaction time. In contrast with the dynamics depicted in Fig.~(\ref{fig2})(b)), the field evolution of  Fig.~(\ref{fig3})(b)) shows that the population of the edge mode with the phase opposite to other three modes slowly decays, causing the field localization in these modes. 

Also, just like for the other discussed cases,  eventually with increasing interaction time the transport  comes to the usual diffusive regime, as is evidenced by the solid and dotted lines.

\section{Discussion of the regimes} 

Here we explain why just the choice of phases of the initial state deeply affects the interference of waveguide modes and leads to so wide a variety of transport regimes. The dissipative dynamics described by Eq.(\ref{lind2}) asymptotically projects any initial state (\ref{inst})  on the state with equal absolute values of amplitudes and $\pi$-phase difference between the neighboring modes extended through the whole lattice. So, the considered states with equal amplitudes and different phases can have very different overlap with the stationary state in dependence on the phase choice. Intuitively, when some initial state turns to zero in the right-hand side of Eq.(\ref{eqg1}) for some group of $\{j,k\}$, one should expect rather slow initial dynamics of the corresponding correlation functions, and \textit{vice versa}. The combination of such "slow" and "fast" regions produces the structuring seen in Figs.~\ref{fig2}(b) and \ref{fig3}(b).

\subsection{Continuous approximation}

This intuition can be further explained and corroborated by a simple qualitative model able to capture interference phenomena leading to the plethora of the observed regimes. To that end  let us introduce the functions $\psi_{j,k}=(-1)^{j+k}g_{j,k}$ and consider the continuous approximation of an infinitely long chain getting from Eq.(\ref{eqg1}) the following 
2D heat-transfer equation \cite{alma990023865100205776,alma990023849800205776}:
\begin{equation}
    \frac{1}{\gamma}\frac{\partial}{\partial t}\psi(x,y;t)=\frac{\partial^2}{\partial x^2}\psi(x,y;t) +\frac{\partial^2}{\partial y^2}\psi(x,y;t),
\label{conteq1}    
\end{equation}
where $x$ and $y$ are normalized dimensionless spatial coordinates.
Curiously, for the imaginary $\gamma$,  Eq.(\ref{conteq1}) describes   propagation  of the field spatial coherence of diffracting beams  \cite{Chriki:19,Smartsev:20}. Coherent diffusion of a complex-valued fields  described by Eq.(\ref{conteq1}) 
can be found also in other fields of physics, for example, when studying atomic coherence in a gas of diffusing atoms \cite{PhysRevLett.101.043601,PhysRevLett.105.183602}, restricted diffusion in a magnetic field \cite{RevModPhys.79.1077},  or spin-transport in semiconductors \cite{1999Natur}. Also, the continuous approximation of Eq.(\ref{conteq1})  can be used to obtain approximate descriptions of coherent diffusion in finite-sized systems with different boundary conditions.

We solve Eq.(\ref{conteq1}) with the following initial condition: 
\[\psi(x,y;0)=\sum\limits_{\forall jk} \phi_{jk}\delta(x-x_j)\delta(y-x_k),\] 
where $\phi_{jk}$ describes the initial excitation distribution of waveguides, and $\delta(x)$ is the Dirac delta function. Thus, one can write down the standard solution for the continuous probability distribution \cite{alma990023849800205776}:
\begin{equation}
 p(x,t)=\sum\limits_{\forall j,k}\frac{c_{jk}(t)}{\sqrt{2\pi{\gamma}t}} \exp{\left\{-\frac{(x-(x_j+x_k)/2)^2}{2{\gamma}t}\right\}}
    \label{sol1}
\end{equation}
where $p(x,t)=\psi(x,x;t)$ and
\begin{equation}
c_{jk}(t)=\frac{\phi_{jk}}{\sqrt{8\pi{\gamma}t}}\exp{\left\{-\frac{(x_k-x_j)^2}{8{\gamma}t}\right\}}.
\label{cjk}
\end{equation}
Equations (\ref{sol1}) and (\ref{cjk}) lead to the following solution for the continuous version of the variance (\ref{var1}): 
\begin{equation}
{\bar\Delta}^2(t)={\gamma}t+D(t),
\label{varcon}
\end{equation}
which is the sum of the classical diffusive part
and the interference part appearing due to coherence of the initial state
\begin{equation}
D(t)=\frac{1}{4{\bar p}_{tot}(t)}{\sum\limits_{\forall j,k}c_{jk}(t)(x_j+x_k)^2} - ({\bar x}(t))^2,
\label{varcoh}
\end{equation}
where the average displacement is 
\begin{eqnarray}
\label{xbar}
{\bar x}(t)=\frac{1}{2{\bar p}_{tot}(t)}
{\sum\limits_{\forall j,k}c_{jk}(t)(x_j+x_k)}
\end{eqnarray}
and the total probability is
\begin{equation}
{\bar p}_{tot}={\sum\limits_{\forall j,k}c_{jk}}.
    \label{ptotcon}
\end{equation}

%\begin{multicols}{2}
\begin{figure*}[htb]
\begin{center}
\includegraphics[width=0.85\textwidth]{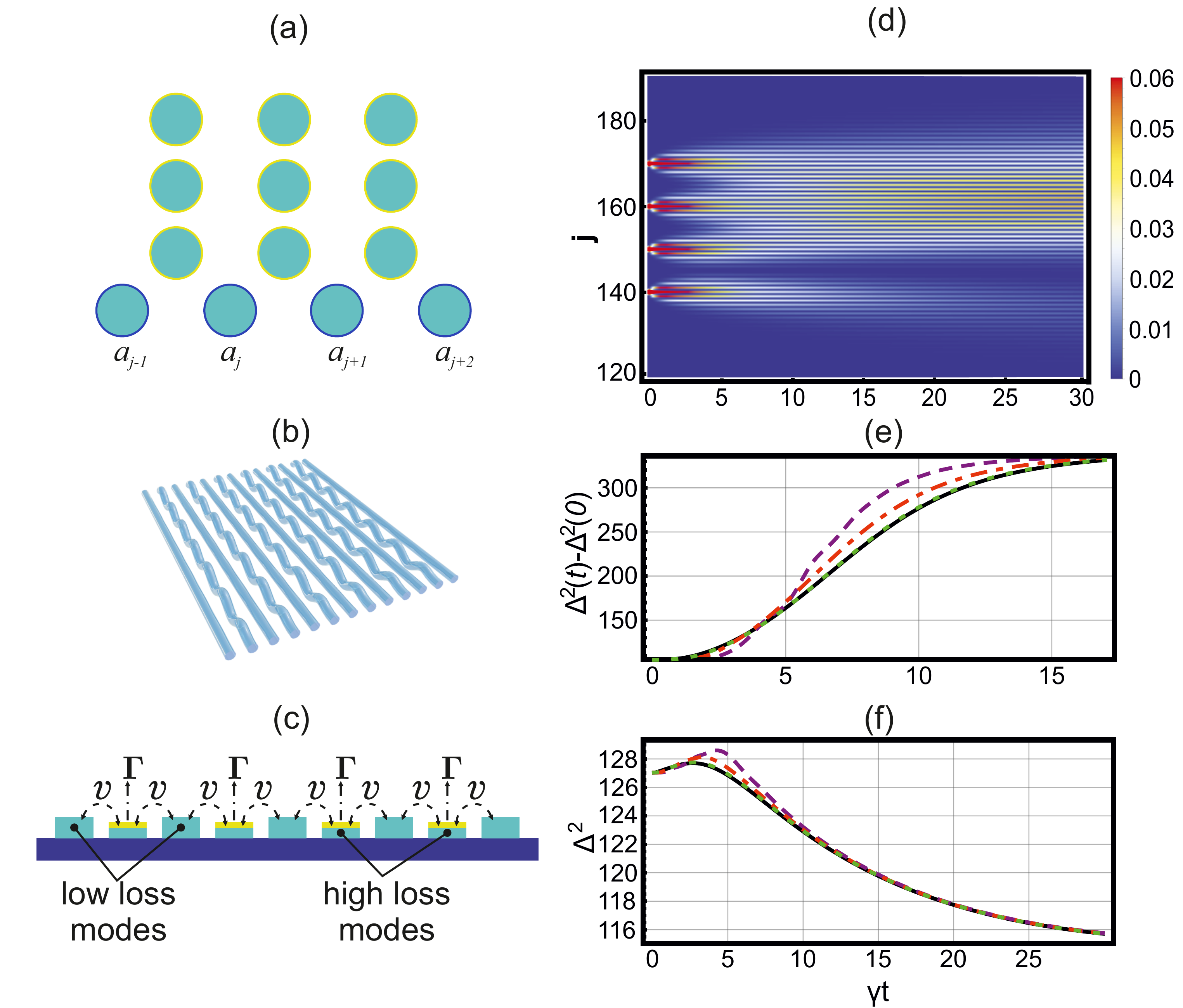}
\end{center}
\caption{ (a) The scheme of coupled single-mode waveguides with designed loss  through "tails" of coupled waveguides as was experimentally realized in Ref.\cite{Mukherjee2017DissipativelyCW}. (b) The scheme of coupled single-mode waveguides with designed loss  through "wavy" modulation of every second waveguide as was experimentally realized in Refs.\cite{eichel,Eichelkraut:14}. (c) The scheme with planar integrated waveguides with induced additional loss in every second waveguides as it is described in Ref.\cite{myarxiv2022}. 
(d) Distribution of the population $p_j$ for all the waveguides including lossy ones and for the initial state  corresponding to the dashed line in Fig.~\ref{fig3}(a).
(e) Variance for the states given by the solution of Eq.(\ref{lind3}) for the initial state of the $a_j$ modes describing by (\ref{inst},\ref{supes1}), with the amplitude $\alpha = 1$, the signature $\{-,+,-,+,-,+\}$ and $m=3$. (f) Variance for the states given by the solution of Eq.(\ref{lind3}) for the initial state of the $a_j$ modes describing by (\ref{inst}), with 4 initially excited modes with numbers $k=n_0-2m,n_0,n_0+2m,n_0+4m$, with the amplitude $\alpha = 1$, the signature $\{+,-,-,-\}$ and $m=5$. For both the panels (e,f) the initial states of the modes $b_j$ are the vacuum and other parameters are as for Fig.(1)(d). Dash-dotted, dashed and dotted lines correspond to $v/\Gamma=0.43,0.25,0.083$. The solid line in panel (e) replicates the dash-dotted line in Fig.~\ref{fig2}(a), and the dashed line in panel (f) replicates  the dashed line in Fig.~\ref{fig3}(a).}
\label{fig4}
\end{figure*}
%\end{multicols}

\subsection{Limitations and results}

Of course, one can hardly expect that the simple  model (\ref{conteq1})  would be able to reproduce precisely all the features of the continuous model. This is obvious even from the  variance dynamics of the initial single-mode excited state shown in Fig.~\ref{fig1}(c). For this case the continuous approximation (\ref{varcon})  gives just a straight line  ignoring the transition stage and resulting in higher values of the variance. Figure ~\ref{newfig} demonstrates that larger distance between the initially excited waveguides improves the validity of the approximation.  Figure ~\ref{newfig}(a) shows that the  variance for the exact solution of Eq.(\ref{lind2})    and for the approximate solution given by Eqs.(\ref{varcon},\ref{varcoh}) become close even for times less than $1/\gamma$ when the distance between the initially excited waveguides grow. However, adding two excited state close to each other to a well-extended state spoils the approximation for small times (see the upper solid and dashed lines in Fig.~\ref{newfig} (a); these lines correspond  to four initially excited modes, $n_0-2,n_0,n_0+1,n_0+3$, with the signature $\{+,-,+,-\}$; the solid lines correspond to the exact solution of Eq.(\ref{lind2}); the dashed lines correspond to the approximation given by Eqs.(\ref{varcon},\ref{varcoh})). Figure ~\ref{newfig} (b) shows a relative difference of variance for the exact and approximate solutions, $\delta_{\Delta^2}=|\Delta^2_{exact}(t)-\Delta^2_{approx}(t)|/\Delta^2_{exact}(t)$. One can see that indeed distancing initially excited waveguides from one another reduces the relative error of the approximation even for the stage $\gamma t < 1$. 
As Fig.~\ref{newfig}(b) shows, for the states illustrated in Fig.~\ref{fig1}(d) the error of the approximation is small.

So, Eqs. (\ref{sol1}-\ref{ptotcon}) can indeed lead to a simple explanation of the observed regimes discussed above. First of all, it is easy to see that deviations from the diffusive regime are indeed due to coherence of the initial state. In the absence of the coherence, i.e.,   when $\phi_{jk}\propto \delta_{jk}$, the term $D(t)$ is constant and is zero for just one initially excited waveguide. This term is also constant  asymptotically (for $\gamma t\rightarrow \infty$), if the sum of the initial state coherences is not zero $s_{coh}=\sum\limits_{\forall j,k} \phi_{jk}\neq 0$. 

It was already shown in Ref.\cite{Mogilevtsev_2015} that having $s_{coh}=0$ can lead to a number of quite nontrivial consequences. In particular, the choice of the initial state can lead to different kinds of asymptotic polynomial decay of $p_{tot}(t)$. Formally, it is quite easy to see from the long-time dynamics corresponding to $(x_k-x_j)^2/{8{\gamma}t}\rightarrow 0$, when  the function $\sqrt{{\gamma}t}c_{jk}(t)$ can be approximated by the finite number of terms of the  series
\begin{equation}
\sqrt{{\gamma}t}c_{jk}(t)\propto \sum\limits_{n=0}\frac{\phi_{jk}}{n!}\left[-\frac{(x_k-x_j)^2}{8{\gamma}t}\right]^n .
    \label{series}
\end{equation}

As it follows from Eqs. (\ref{sol1}-\ref{series}), for the initial state (\ref{inst}) having $s_{coh}=0$  and 
$\sum\limits_{\forall j,k}\phi_{jk}x_kx_j\neq 0$
leads to $\sqrt{{\gamma}t}{\bar p}_{tot}(t\rightarrow \infty)\propto (\gamma t)^{-1}$ and $D(t\rightarrow \infty)=2{\gamma}t$ giving rise to the 3 times larger asymptotic diffusion rate observed in  Figs.~\ref{fig1}, \ref{fig2}. 

One can see also from Eqs. (\ref{sol1}-\ref{series}) that different polynomial decay laws of $p_{tot}(t)$ can be achieved by the choice of the initial state (\ref{inst}) zeroing first terms of the series (\ref{series}) representing  $p_{tot}(t)$. Such a choice of the initial state (\ref{inst}) will not change the asymptotically linear behavior of $D(t\rightarrow \infty)$. However, it might affect the dynamics at earlier times. Exactly this is shown in Figs.~\ref{fig2},\ref{fig3}. Interference cancels lower-order terms making faster terms to contribute and thus creating superballistic transport dynamics. In the localization case, interference can make $D(t)$ negative, as is seen in Fig.~\ref{fig3}.  

Finally, Eqs. (\ref{sol1}-\ref{series}) clearly show the timescaling demonstrated in  Figs.~\ref{fig2},\ref{fig3}. Indeed, the time dependence of $D(t)$ is only through the terms $(x_k-x_j)^2/{\gamma}t$. Having increased every distance $x_k-x_j$ $m$ times is equivalent to decreasing the dissipative coupling rate, ${\gamma}$, $m^2$ times.

However, one should emphasize that  the scaling behavior of the variance is not reduced to a simple, albeit nonlinear, stretching along the time axis. As is shown in Figs.~\ref{fig2}(a),\ref{fig3}(a), with increasing the distance between the initially excited waveguides, not only are the regions of superdiffusive and subdiffusive behavior extending, but the values of variance reached before coming to the diffusive regime are different, too. 

\section{Estimations for potential experiment} 

The considered system of dissipatively coupled modes (\ref{lind3})   can be realized and is already realized as an array of unitary coupled single-mode waveguides where every second waveguide has an enhanced designed loss.  Such systems were made by laser-writing waveguides in glass \cite{eichel,grafe,loss1,Mukherjee2017DissipativelyCW}. In particular, designed loss in Ref.\cite{eichel,grafe} was realized by making "wavy" modulated waveguides ( see Fig.~\ref{fig4}(a)). Of course, making waveguides wavy means that both loss and coupling to the neighboring straight waveguides varies. However, the interaction length required to observe the transition between the ballistic and diffusive dynamics (several centimeters)  far exceeds the modulated waveguide oscillation length (around a millimeter), allowing one to describe the system with an average coupling coefficient of about 0.5 cm$^{-1}$ and the designed loss with a rate of about 0.5 - 1 cm$^{-1}$. Notice that for the borosilicate glass and the used 633 nm wavelength  the passive loss rate on the written straight waveguides was about 0.1 cm$^{-1}$ Ref. \cite{Eichelkraut:14}.  In Ref.\cite{Mukherjee2017DissipativelyCW} the designed loss was realized by coupling every second waveguide to "tails" of waveguides ( see Fig.~\ref{fig4}(b)) . There the two-waveguide equalization and the state transfer across several dissipatively coupled waveguides were experimentally demonstrated with cw coherent excitation using 700 - 800 nm light. There also the coupling strength and the designed loss were of an order of magnitude higher than the passive waveguide loss. 

More versatile and compact structures can be realized on integrated photonic platforms, such as the InP platform \cite{jstqe-augu18,myarxiv2022}. The latter realization is depicted in Fig.~\ref{fig4}(c), where the coupling between waveguides is with the coupling constant $v$, and every second waveguide is subject to the designed loss with the rate $\Gamma$. As is discussed in Ref.\cite{myarxiv2022}, realization  on an integrated photonic platform can be more versatile than waveguide-writing schemes due to greater variability of ways and degrees of modification of the designed loss. One can  achieve up to 2 orders of magnitude larger coupling and designed loss than in the glass-written structures and can observe the discussed transport regimes on a few millimeter scale with cw light.

Notice that the system (\ref{lind3}) discussed in Ref.\cite{eichel} for a single initially excited waveguide demonstrated the appearance  of the diffusive regime typical for the dissipatively coupled bosonic chains discussed here with the transition time of about $1/\gamma$ even not being in the dissipative coupling limit. 
Figures ~\ref{fig4}(e,f) show how for the realistic structures of Figs.~\ref{fig4}(a-c) the dissipative coupling limit is approached for the superballistic regime (Fig.~\ref{fig4}(e)) and the localization (Fig.~\ref{fig4}(f)) regime with diminishing the ratio  $v/\Gamma$  while keeping $\gamma$ constant. Dashed, dotted and dash-dotted lines in Figs.~\ref{fig4}(e) were obtained for the initial state (\ref{inst}) with excitation as it was for the dash-dotted line in Fig.~\ref{fig2}(a).  Dashed, dotted and dash-dotted lines in Figs.~\ref{fig4}(f) were obtained for the initial state (\ref{inst}) with excitation as it was for the dashed line in Fig.~\ref{fig3}(a). For both Figs.~\ref{fig4}(e,f) dash-dotted, dashed and dotted lines correspond to $v/\Gamma=0.43,0.25,0.083$. All the $b_j$ modes were supposed initially to be in the vacuum states. Solid lines in Figs.~\ref{fig4}(e,f) show corresponding solutions of Eq.(\ref{lind2}).

It can be seen in Figs.~\ref{fig4}(e,f) that the dynamics typical for the limit of large loss is actually quite easily reached in practice.  Already for $v\approx 0.25 \Gamma$ one has a good approximation of the dissipative coupling regime, and $v\approx 0.08\Gamma$ gives results practically coinciding with ones obtained from Eq.(\ref{lind2}). Figure~\ref{fig4}(d) shows the normalized population distribution $p_j$  in all the waveguides including lossy ones for the localization regime corresponding to the dotted line in Fig.~\ref{fig4}(f). As it should be expected in the regime of the dissipative coupling, population in the lossy waveguides drops very quickly (on the scale of $\Gamma t\sim1$), reproducing the picture seen in Fig.~\ref{fig3} (b).

\section{Conclusions} 

Here we have demonstrated that perfectly periodic networks of dissipatively coupled bosonic modes can exhibit a variety of transport regimes in dependence on the initial excitation of the modes. The discussed regimes can be observed in tight-bonding 1D systems of coupled single-mode waveguides feasible in different photonic platforms. Such systems were already realized by laser writing in balk glass, and can be also realized with integrated photonic platforms.  Our observations are important in view of the possible applications of these networks of dissipatively coupled modes. For  functioning optical equalizers and beam splitters, distributors and nonreciprocal optical circuits need input fields to spread across a network \cite{Mukherjee2017DissipativelyCW,myarxiv2022}. Studies of on the field propagation in the networks can allow for optimization of propagation time and regimes and, in particular, for minimization of the device length.  

We have shown that coherent interference effects can lead to a superballistic transport regime or to localization in any predefined interaction time interval. This interval can be extended by changing the initial state. We have described the way of doing that and derived a simple analytic model to explain such scaling behavior. This model also explains the appearance of both localization and superballistic regimes and demonstrates the nature of different possible rates of the asymptotic transport regime. It is always diffusive. However, for certain initial states the rate of diffusion can be three times more than for others. 

Notice that here we have pursued only the task of demonstrating possibilities offered by interference occurring in the coherent diffusive quantum walk of amplitudes. We have discussed transport regimes for different classical factorized states of the bosonic modes. We have not exploited correlated states and propagation of correlation and nonclassicality through our network. To provide a hint toward possible nontrivial effects connected with quantum correlation transport, one can notice that the finite system described by Eq.(\ref{lind2}) has entangled stationary states \cite{Mogilevtsev_2015,Mukherjee2017DissipativelyCW}. Also, the dynamics of the $n$-th order correlation function describes coherent diffusion in $n$-dimensions.

The authors (D.M and I.P) gratefully acknowledge support from  the BRFFI Projects No. F21KOP-002 and No. F22B-008.  G.S. acknowledges support from the Horizon 2020 project TERASSE 823878 and from the NATO project NATO SPS - G5860.

%\bibliography{refpre}

\begin{thebibliography}{58}
\providecommand{\natexlab}[1]{#1}
\providecommand{\url}[1]{\texttt{#1}}
\expandafter\ifx\csname urlstyle\endcsname\relax
  \providecommand{\doi}[1]{doi: #1}\else
  \providecommand{\doi}{doi: \begingroup \urlstyle{rm}\Url}\fi

\bibitem[{Davis} et~al.(1996){Davis}, {Miura}, {Sugimoto}, and
  {Hirao}]{davis1996}
K.~M. {Davis}, K.~{Miura}, Naoki {Sugimoto}, and Kazuyuki {Hirao},
\newblock {Writing waveguides in glass with a femtosecond laser},
\newblock \emph{Optics Letters} \textbf{21}, 1729
  (1996).

\bibitem[Szameit and Nolte(2010)]{Szameit_2010}
A.~Szameit and S.~Nolte,
\newblock Discrete optics in femtosecond-laser-written photonic structures,
\newblock \emph{Journal of Physics B: Atomic, Molecular and Optical Physics}
  \textbf{43}, 163001 (2010).

\bibitem[{Koll{\'a}r} et~al.(2019){Koll{\'a}r}, {Fitzpatrick}, and
  {Houck}]{2019Natur}
A.~J. {Koll{\'a}r}, M.~{Fitzpatrick}, and A.~A. {Houck}.
\newblock {Hyperbolic lattices in circuit quantum electrodynamics}, \textit{Nature}
\textbf {571}, 45 (2019).

\bibitem[Sheng et~al.(2022)Sheng, Wang, Chang, Wang, Lu, Yang, Zhu, Jin, and
  Liu]{sheng}
Chong Sheng, Yao Wang, Yijun Chang, Huiming Wang, Yongheng Lu, Yingyue Yang,
  Shining Zhu, Xianmin Jin, and Hui Liu,
\newblock Bound vortex light in an emulated topological defect in photonic
  lattices,
\newblock \emph{Light: Science $\&$ Applications} \textbf{11}, 243 (2022).

\bibitem[Dreisow et~al.(2012)Dreisow, Longhi, Nolte, T\"unnermann, and
  Szameit]{PhysRevLett.109.110401}
F.~Dreisow, S.~Longhi, S.~Nolte, A.~T\"unnermann, and A.~Szameit,
\newblock Vacuum instability and pair production in an optical setting,
\newblock \emph{Phys. Rev. Lett.} \textbf{109}, 110401 (2012).

\bibitem[Vicencio et~al.(2015)Vicencio, Cantillano, Morales-Inostroza, Real,
  Mej\'{\i}a-Cort\'es, Weimann, Szameit, and Molina]{PhysRevLett.114.245503}
R.~A. Vicencio, C.~Cantillano, L.~Morales-Inostroza, B.~Real,
  C.~Mej\'{\i}a-Cort\'es, S.~Weimann, A.~Szameit, and M.~I. Molina,
\newblock Observation of localized states in Lieb photonic lattices.
\newblock \emph{Phys. Rev. Lett.} \textbf{114}, 245503 (2015).

\bibitem[Mukherjee et~al.(2015)Mukherjee, Spracklen, Choudhury, Goldman,
  \"Ohberg, Andersson, and Thomson]{PhysRevLett.114.245504}
S.~Mukherjee, A.~Spracklen, D.~Choudhury, N.~Goldman, P.~\"Ohberg,
  E.~Andersson, and R.~R. Thomson,
\newblock Observation of a localized flat-band state in a photonic Lieb
  lattice,
\newblock \emph{Phys. Rev. Lett.} \textbf{114}, 245504 (2015).

\bibitem[{Wang} et~al.(2020){Wang}, {Zheng}, {Chen}, {Huang}, {Kartashov},
  {Torner}, {Konotop}, and {Ye}]{wang2020}
Peng {Wang}, Yuanlin {Zheng}, Xianfeng {Chen}, Changming {Huang}, Y.~V.
  {Kartashov}, Lluis {Torner}, V.~V. {Konotop}, and Fangwei {Ye},
\newblock {Localization and delocalization of light in photonic Moir{\'e}
  lattices},
\newblock \emph{Nature} \textbf{577}, 42 (2020).

\bibitem[{Longhi}(2014)]{longhi2014}
S.~{Longhi},
\newblock {Aharonov-Bohm photonic cages in waveguide and coupled resonator
  lattices by synthetic magnetic fields},
\newblock \emph{Optics Letters} \textbf{39}, 5892 (2014).

\bibitem[Mukherjee et~al.(2018)Mukherjee, Di~Liberto, \"Ohberg, Thomson, and
  Goldman]{PhysRevLett.121.075502}
S.~Mukherjee, M.~Di~Liberto, P.~\"Ohberg, R.~R. Thomson, and N.~Goldman,
\newblock Experimental observation of aharonov-bohm cages in photonic lattices,
\newblock \emph{Phys. Rev. Lett.} \textbf{121} 075502 (2018).

\bibitem[{Kremer} et~al.(2020){Kremer}, {Petrides}, {Meyer}, {Heinrich},
  {Zilberberg}, and {Szameit}]{kremer}
M.~{Kremer}, I.~{Petrides}, E.~{Meyer}, M.~{Heinrich}, O.~{Zilberberg}, and
  A.~{Szameit},
\newblock {A square-root topological insulator with non-quantized indices
  realized with photonic Aharonov-Bohm cages},
\newblock \emph{Nature Communications} \textbf{11}, 907 (2020).

\bibitem[Cerjan et~al.(2020)Cerjan, J\"urgensen, Benalcazar, Mukherjee, and
  Rechtsman]{PhysRevLett.125.213901}
A.~Cerjan, M.~J\"urgensen, W.~A. Benalcazar, S.~Mukherjee, and M.~C. Rechtsman,
\newblock Observation of a higher-order topological bound state in the
  continuum,
\newblock \emph{Phys. Rev. Lett.} \textbf{125} 213901 (2020).

\bibitem[{Rechtsman} et~al.(2013){Rechtsman}, {Zeuner}, {Plotnik}, {Lumer},
  {Podolsky}, {Dreisow}, {Nolte}, {Segev}, and {Szameit}]{rechtsman}
M.~C. {Rechtsman}, J.~M. {Zeuner}, Y.~{Plotnik}, Y.~{Lumer}, D.~{Podolsky},
  F.~{Dreisow}, S.~{Nolte}, M.~{Segev}, and A.~{Szameit},
\newblock {Photonic Floquet topological insulators},
\newblock \emph{Nature} \textbf{496}, 196 (2013).

\bibitem[Yuan et~al.(2018)Yuan, Lin, Xiao, and Fan]{Yuan18}
Luqi Yuan, Qian Lin, Meng Xiao, and Shanhui Fan,
\newblock Synthetic dimension in photonics,
\newblock \emph{Optica} \textbf{5}, 1396 (2018).

\bibitem[Buljan et~al.(2022)Buljan, Jukić, and Chen]{loss}
H.~Buljan, D.~Jukić, and Zhigang Chen,
\newblock Loss leads the way to utopia, 
\newblock \emph{Nature Physics} \textbf{18}, 371 (2022).

\bibitem[Metelmann and Clerk(2015)]{PhysRevX.5.021025}
A.~Metelmann and A.~A. Clerk,
\newblock Nonreciprocal photon transmission and amplification via reservoir
  engineering,
\newblock \emph{Phys. Rev. X} \textbf{5}, 021025 (2015).

\bibitem[Arenz and Metelmann(2020)]{arenz}
C.~Arenz and A.~Metelmann,
\newblock Emerging unitary evolutions in dissipatively coupled systems,
\newblock \emph{Phys. Rev. A} \textbf{101}, 022101 (2020).

\bibitem[Huang et~al.(2021)Huang, Lu, Liang, Tao, and Liu]{Huang2021}
Xinyao Huang, Cuicui Lu, Chao Liang, Honggeng Tao, and Y.~Liu,
\newblock Loss-induced nonreciprocity,
\newblock \emph{Light, Science \& Applications} \textbf{10}, 30 (2021).

\bibitem[Peshko et~al.(2022)Peshko, Pustakhod, and Mogilevtsev]{myarxiv2022}
I.~Peshko, D.~Pustakhod, and D.~Mogilevtsev,
\newblock Breaking reciprocity by designed loss,
\newblock \emph{J. Opt. Soc. Am. B} \textbf{39}, 1926
  (2022).

\bibitem[Longhi(2020)]{longhi2020}
S.~Longhi,
\newblock Unraveling the non-Hermitian skin effect in dissipative systems,
\newblock \emph{Phys. Rev. B}, \textbf{102}, 201103 (2020).

\bibitem[Carvalho et~al.(2001)Carvalho, Milman, de~Matos~Filho, and
  Davidovich]{PhysRevLett.86.4988}
A.~R.~R. Carvalho, P.~Milman, R.~L. de~Matos~Filho, and L.~Davidovich,
\newblock Decoherence, pointer engineering, and quantum state protection,
\newblock \emph{Phys. Rev. Lett.} \textbf{86}, 4988 (2001).

\bibitem[Thornton et~al.(2019)Thornton, Sakovich, Mikhalychev, Ferrer, de~la
  Hoz, Korolkova, and Mogilevtsev]{PhysRevApplied.12.064051}
M.~Thornton, A.~Sakovich, A.~Mikhalychev, J.~D. Ferrer, P.~de~la Hoz,
  N.~Korolkova, and D.~Mogilevtsev,
\newblock Coherent diffusive photon gun for generating nonclassical states,
\newblock \emph{Phys. Rev. Applied} \textbf{12}, 064051  (2019).

\bibitem[Mukherjee et~al.(2017)Mukherjee, Mogilevtsev, Slepyan, Doherty,
  Thomson, and Korolkova]{Mukherjee2017DissipativelyCW}
S.~Mukherjee, D.~Mogilevtsev, G.~Slepyan, T.~H. Doherty, R.~Thomson, and
  N.~Korolkova,
\newblock Dissipatively coupled waveguide networks for coherent diffusive
  photonics,
\newblock \emph{Nature Communications} \textbf{8}, 1909 (2017).

\bibitem[{Biggerstaff} et~al.(2016){Biggerstaff}, {Heilmann}, {Zecevik},
  {Gr{\"a}fe}, {Broome}, {Fedrizzi}, {Nolte}, {Szameit}, {White}, and
  {Kassal}]{bigger}
D.~N. {Biggerstaff}, R.~{Heilmann}, A.~A. {Zecevik}, M.~{Gr{\"a}fe}, M.~A.
  {Broome}, A.~{Fedrizzi}, S.~{Nolte}, A.~{Szameit}, A.~G. {White}, and
  I.~{Kassal},
\newblock {Enhancing coherent transport in a photonic network using
  controllable decoherence},
\newblock \emph{Nature Communications} \textbf{7}, 11282 (2016).

\bibitem[Eichelkraut et~al.(2013)Eichelkraut, Heilmann, Weimann, Stützer,
  Dreisow, Christodoulides, Nolte, and Szameit]{eichel}
T.~Eichelkraut, R~Heilmann, S.~Weimann, S.~Stützer, F~Dreisow,
  D~Christodoulides, S.~Nolte, and A.~Szameit,
\newblock Mobility transition from ballistic to diffusive transport in
  non-Hermitian lattices,
\newblock \emph{Nature Communications} \textbf{4}, 2533 (2013).

\bibitem[Gräfe et~al.(2016)Gräfe, Heilmann, Lebugle, Guzman-Silva,
  Perez-Leija, and Szameit]{grafe}
M.~Gräfe, R.~Heilmann, M.~Lebugle, D.~Guzman-Silva, A.~Perez-Leija, and
  A.~Szameit,
\newblock Integrated photonic quantum walks,
\newblock \emph{Journal of Optics} \textbf{18}, 103002 (2016).

\bibitem[Kendon(2007)]{kendon}
V. Kendon,
\newblock Decoherence in quantum walks – a review,
\newblock \emph{Mathematical Structures in Computer Science} \textbf{17}, 1169 (2007).

\bibitem[Romanelli et~al.(2005)Romanelli, Siri, Abal, Auyuanet, and
  Donangelo]{romanelli}
A.~Romanelli, R.~Siri, G.~Abal, A.~Auyuanet, and R.~Donangelo,
\newblock Decoherence in the quantum walk on the line,
\newblock \emph{Phys. A (Amsterdam, Neth.)}
  \textbf{347}, 137 (2005).


\bibitem[Golshani et~al.(2014{\natexlab{a}})Golshani, Weimann, Jafari, Nezhad,
  Langari, Bahrampour, Eichelkraut, Mahdavi, and Szameit]{loss1}
M.~Golshani, S.~Weimann, Kh. Jafari, M.~K. Nezhad, A.~Langari, A.~R.
  Bahrampour, T.~Eichelkraut, S.~M. Mahdavi, and A.~Szameit,
\newblock Impact of loss on the wave dynamics in photonic waveguide lattices,
\newblock \emph{Phys. Rev. Lett.} \textbf{113}, 123903
  (2014).



\bibitem[Aharonov et~al.(1993)Aharonov, Davidovich, and Zagury]{aharon93}
Y.~Aharonov, L.~Davidovich, and N.~Zagury,
\newblock Quantum random walks,
\newblock \emph{Phys. Rev. A}, \textbf{48}, 1687 (1993).

\bibitem[Kempe(2003)]{kempe}
J.~Kempe,
\newblock Quantum random walks: An introductory overview,
\newblock \emph{Contemporary Physics} \textbf{44}, 307
  (2003).

\bibitem[Venegas-Andraca(2012)]{venegas}
S.~Venegas-Andraca,
\newblock Quantum walks: A comprehensive review,
\newblock \emph{Quantum Inf. Proc.} \textbf{11}, 01 (2012).

\bibitem[Farhi and Gutmann(1998)]{farhi}
E.~Farhi and S.~Gutmann,
\newblock Quantum computation and decision trees,
\newblock \emph{Phys. Rev. A} \textbf{58}, 915 (1998).

\bibitem[Mülken and Blumen(2011)]{MULKEN201137}
O.~Mülken and A.~Blumen,
\newblock Continuous-time quantum walks: Models for coherent transport on
  complex networks,
\newblock \emph{Physics Reports} \textbf{502}, 37 (2011).


\bibitem[Perets et~al.(2008)Perets, Lahini, Pozzi, Sorel, Morandotti, and
  Silberberg]{PhysRevLett.100.170506}
H.~B. Perets, Y.~Lahini, F.~Pozzi, M.~Sorel, R.~Morandotti, and Y.~Silberberg,
\newblock Realization of quantum walks with negligible decoherence in waveguide
  lattices,
\newblock \emph{Phys. Rev. Lett.} \textbf{100}, 170506 (2008).

\bibitem[M\"ulken et~al.(2007)M\"ulken, Blumen, Amthor, Giese, Reetz-Lamour,
  and Weidem\"uller]{PhysRevLett.99.090601}
O.~M\"ulken, A.~Blumen, T.~Amthor, C.~Giese, M.~Reetz-Lamour, and
  M.~Weidem\"uller,
\newblock Survival probabilities in coherent exciton transfer with trapping,
\newblock \emph{Phys. Rev. Lett.} \textbf{99}, 090601 (2007).

\bibitem[Rebentrost et~al.(2009)Rebentrost, Mohseni, and Aspuru-Guzik]{reben}
P.~Rebentrost, M.~Mohseni, and A.~Aspuru-Guzik,
\newblock Role of quantum coherence and environmental fluctuations in
  chromophoric energy transport,
\newblock \emph{The Journal of Physical Chemistry B} \textbf{113}, 9942 (2009).

\bibitem[Chia et~al.(2016)Chia, Tan, Pawela, Kurzy\ifmmode~\acute{n}\else
  \'{n}\fi{}ski, Paterek, and Kaszlikowski]{PhysRevE.93.032407}
A.~Chia, K.~C. Tan, \L{}. Pawela, P.~Kurzy\ifmmode~\acute{n}\else
  \'{n}\fi{}ski, T.~Paterek, and D.~Kaszlikowski,
\newblock Coherent chemical kinetics as quantum walks. I. Reaction operators  for radical pairs,
\newblock \emph{Phys. Rev. E}, \textbf{93}, 032407 (2016).

\bibitem[Levi et~al.(2011)Levi, Rechtsman, Freedman, Schwartz, Manela, and
  Segev]{liad}
L.~Levi, M.~Rechtsman, B.~Freedman, T.~Schwartz, O.~Manela, and M.~Segev,
\newblock Disorder-enhanced transport in photonic quasicrystals,
\newblock \emph{Science} \textbf{332}, 1541 (2011).

\bibitem[{Levi} et~al.(2012){Levi}, {Krivolapov}, {Fishman}, and
  {Segev}]{levi2012}
L.~{Levi}, Y.~{Krivolapov}, S.~{Fishman}, and M.~{Segev},
\newblock {Hyper-transport of light and stochastic acceleration by evolving
  disorder},
\newblock \emph{Nature Physics} \textbf{8}, 912 (2012).

\bibitem[Zhang et~al.(2012)Zhang, Tong, Gong, and Li]{PhysRevLett.108.070603}
Zhenjun Zhang, Peiqing Tong, Jiangbin Gong, and Baowen Li,
\newblock Quantum hyperdiffusion in one-dimensional tight-binding lattices,
\newblock \emph{Phys. Rev. Lett.} \textbf{108}, 070603 (2012).

\bibitem[Pikovsky and Shepelyansky(2008)]{PhysRevLett.100.094101}
A.~S. Pikovsky and D.~L. Shepelyansky,
\newblock Destruction of Anderson localization by a weak nonlinearity,
\newblock \emph{Phys. Rev. Lett.} \textbf{100}, 094101 (2008).

\bibitem[Held et~al.(2022)Held, Engelkemeier, De, Barkhofen, Sperling, and
  Silberhorn]{PhysRevA.105.042210}
P.~Held, M.~Engelkemeier, S.~De, S.~Barkhofen, J.~Sperling, and Ch. Silberhorn,
\newblock Driven Gaussian quantum walks,
\newblock \emph{Phys. Rev. A} \textbf{105}, 042210 (2022).

\bibitem[Schreiber et~al.(2011)Schreiber, Cassemiro,
  Poto\ifmmode~\check{c}\else \v{c}\fi{}ek, G\'abris, Jex, and
  Silberhorn]{PhysRevLett.106.180403}
A.~Schreiber, K.~N. Cassemiro, V.~Poto\ifmmode~\check{c}\else \v{c}\fi{}ek,
  A.~G\'abris, I.~Jex, and Ch. Silberhorn,
\newblock Decoherence and disorder in quantum walks: From ballistic spread to
  localization,
\newblock \emph{Phys. Rev. Lett.} \textbf{106}, 180403 (2011).

\bibitem[Sch\"utz and Trimper(2004)]{PhysRevE.70.045101}
G.~M. Sch\"utz and S.~Trimper,
\newblock Elephants can always remember: Exact long-range memory effects in a
  non-Markovian random walk,
\newblock \emph{Phys. Rev. E} \textbf{70}, 045101 (2004).

\bibitem[Mogilevtsev et~al.(2015)Mogilevtsev, Slepyan, Garusov, Kilin, and
  Korolkova]{Mogilevtsev_2015}
D.~Mogilevtsev, G.~Ya. Slepyan, E.~Garusov, S.~Ya. Kilin, and N.~Korolkova,
\newblock Quantum tight-binding chains with dissipative coupling,
\newblock \emph{New Journal of Physics} \textbf{17}, 043065 (2015).

\bibitem[Mogilevtsev and Shchesnovich(2010)]{Mogilevtsev:10}
D.~Mogilevtsev and V.~S. Shchesnovich,
\newblock Single-photon generation by correlated loss in a three-core optical
  fiber,
\newblock \emph{Optics Letters} \textbf{35}, 3375 (2010).

\bibitem[Lawler(2010)]{alma990023849800205776}
G.~F. Lawler,
\newblock \emph{Random Walk and the Heat Equation},
\newblock Student Mathematical Library Vol. \textbf{55} (American Mathematical Society,
  Providence, RI, 2010).


\bibitem[Reif(2009)]{alma990023865100205776}
F.~Reif,
\newblock \emph{Fundamentals of Statistical and Thermal Physics}
\newblock (Waveland, Long Grove, IL, 2009).


\bibitem[Chriki et~al.(2019)Chriki, Smartsev, Eger, Firstenberg, and
  Davidson]{Chriki:19}
R.~Chriki, S.~Smartsev, D.~Eger, O.~Firstenberg, and N.~Davidson,
\newblock Coherent diffusion of partial spatial coherence,
\newblock \emph{Optica} \textbf{6}, 1406 (2019).

\bibitem[Smartsev et~al.(2020)Smartsev, Chriki, Eger, Firstenberg, and
  Davidson]{Smartsev:20}
S.~Smartsev, R.~Chriki, D.~Eger, O.~Firstenberg, and N.~Davidson,
\newblock Structured beams invariant to coherent diffusion,
\newblock \emph{Optics Express} \textbf{28}, 33708O  (2020).

\bibitem[Xiao et~al.(2008)Xiao, Klein, Hohensee, Jiang, Phillips, Lukin, and
  Walsworth]{PhysRevLett.101.043601}
Yanhong Xiao, M.~Klein, M.~Hohensee, Liang Jiang, D.~F. Phillips, M.~D. Lukin,
  and Ronald~L. Walsworth,
\newblock Slow light beam splitter,
\newblock \emph{Phys. Rev. Lett.} \textbf{101}, 043601 (2008).

\bibitem[Firstenberg et~al.(2010)Firstenberg, London, Yankelev, Pugatch,
  Shuker, and Davidson]{PhysRevLett.105.183602}
O.~Firstenberg, P.~London, D.~Yankelev, R.~Pugatch, M.~Shuker, and N.~Davidson,
\newblock Self-similar modes of coherent diffusion,
\newblock \emph{Phys. Rev. Lett.} \textbf{105}, 183602 (2010).

\bibitem[Grebenkov(2007)]{RevModPhys.79.1077}
D.~S. Grebenkov,
\newblock NMR survey of reflected Brownian motion,
\newblock \emph{Rev. Mod. Phys.} \textbf{79}, 1077 (2007).

\bibitem[{Kikkawa} and {Awschalom}(1999)]{1999Natur}
J.~M. {Kikkawa} and D.~D. {Awschalom},
\newblock {Lateral drag of spin coherence in gallium arsenide},
\newblock \emph{Nature} \textbf{397}, 139 (1999).

\bibitem[Eichelkraut et~al.(2014)Eichelkraut, Weimann, St\"{u}tzer, Nolte, and
  Szameit]{Eichelkraut:14}
T.~Eichelkraut, S.~Weimann, S.~St\"{u}tzer, S.~Nolte, and A.~Szameit,
\newblock Radiation-loss management in modulated waveguides,
\newblock \emph{Optics Letters} \textbf{39}, 6831 (2014).

\bibitem[Augustin et~al.(2018)Augustin, Santos, den Haan, Kleijn, Thijs,
  Latkowski, Zhao, Yao, Bolk, Ambrosius, Mingaleev, Richter, Bakker, and
  Korthorst]{jstqe-augu18}
L.~M. Augustin, R.~Santos, E.~den Haan, S.~Kleijn, P.~J.~A. Thijs,
  S.~Latkowski, D.~Zhao, W.~Yao, J.~Bolk, H.~Ambrosius, S.~Mingaleev,
  A.~Richter, A.~Bakker, and T.~Korthorst,
\newblock {InP-based generic foundry platform for photonic integrated
  circuits},
\newblock \emph{IEEE J. Sel. Topics in Quantum Electron.} \textbf{24}, 1 (2018).


\end{thebibliography}

\end{document}